\documentclass[aps,prl,twocolumn,showpacs]{revtex4}

\usepackage{amsbsy}
\usepackage{graphicx}

\begin{document}

\title{Shiba multiplets due to ${\rm Mn}$ impurities in ${\rm MgB}_2$}
\author{C\u at\u alin Pa\c scu  Moca$^{1,2,4}$,
Gergely Zar\'and$^{1,3,4}$ Eugene Demler$^{1,3}$ and Boldizs\'ar
Jank\'o$^{1,2}$} \address{ $^1$Materials Science Division, Argonne
National Laboratory, 9700 South Cass Avenue, Argonne, Illinois
60439
\\
$^2$ Department of Physics, University of Notre Dame,  Notre Dame, Indiana, 46556
\\
$^3$ Lyman Physics Laboratory, Harvard University, Cambridge MA
\\
$^4$ Budapest University of Technology and Economics, H-1521 Budapest, Hungary}
\date{\today}
\begin{abstract}
{We study the effect of magnetic ${\rm Mn}$ ions on the two-band
superconductor ${\rm MgB}_2$, and compute both the total and spin
resolved scanning tunneling spectrum in the vicinity of the
magnetic impurity. We show that when the internal structure of the
${\rm Mn}$ ion's $d$-shell is taken into account, multiple Shiba
states appear in the spectrum. These multiplets were missed by
previous calculations based on simplistic models, and their
presence could alter significantly the overall interpretation of
local tunneling spectra for a wide range of superconducting hosts
and magnetic impurities.}
\end{abstract}
\maketitle

The interaction between a single magnetic impurity and the
superconducting host reveals fundamental properties of both the
magnetic ion and the host material. This interaction was first
studied theoretically, within the framework of BCS
superconductivity. During the late sixties Shiba \cite{Shiba}
showed that a magnetic impurity pulls down from the continuum
states a pair of bound states inside the superconducting gap.
Indirect indication for the presence of finite spectral weight
inside the gap of an impure superconductor could be inferred from
global probes of the density of states. However, direct evidence
for the existence of the so-called Shiba states requires an
accurate measurement of the {\em local} density of states near the
impurity. Such a measurement became available only recently by
using high vacuum, low temperature scanning tunneling spectroscopy
(STS). Yazdani and his coworkers imaged\cite{Yazdani} the local
density of states around ${\rm Mn}$ and ${\rm Gd}$ impurities deposited
onto ${\rm Nb}$ single crystals. They found clear evidence for localized
states in the vicinity of the magnetic impurities, in qualitative
agreement with Shiba's original findings, and also with their own
model calculation based on a non-selfconsistent solution to
Bogoliubov-de Gennes equations. Quantitative discrepancies,
however, are also clearly present, especially when comparing the
width and spatial dependence of the resonances to theoretical
expectations. The presence of magnetic impurity induced bound
states in a superconductor was turned around and used, both
theoretically\cite{Scalapino,Flatte,Balatsky} and
experimentally\cite{Pan}, as an investigative tool to probe the
unusual ground state of the cuprate superconductors.

In all the studies carried out so far, the direct comparison
between theory and experiment is hampered by the fact that the
theoretical approaches, following Shiba's original work, use only
a simplified (predominantly classical) model to describe the
magnetic impurity, and assume a single spin one-half electron
channel that couples to the magnetic impurity. Furthermore, the
coupling is assumed to be in the s-wave channel, and spin-orbit
coupling is generally ignored. In reality, magnetic impurities
usually have a much more complicated internal structure
\cite{blandin}: The magnetic moments are usually due to low-lying
and crystal-field split $d$- or $f$-levels with multiple
occupancy. The aim of this paper is to demonstrate that this
internal structure has an impact on the structure of the Shiba
states, and only part of the physics is captured by the simplified
model used before. In particular, {\em multiple channels} of
charge carriers couple to the magnetic impurity through
channel-dependent coupling. The combination of these ingredients
generally lead to the appearance of {\em multiple} pairs of Shiba
states. We compute the spatial and spin structure of the scanning
tunneling microscopy (STM) spectra around the magnetic impurity
and show that these states appear as distinct resonances inside
the superconducting gap, and can be most clearly resolved in spin
resolved STM spectra.

In the following  we illustrate our results on the specific case
of ${\rm Mn}$-doped ${\rm MgB}_2$, but we wish to emphasize that
much of our discussions  carry over to other systems as well, and
that our conclusions are rather general. There are several reasons
to choose the ${\rm Mn-MgB}_2$ system. First, in order to observe
a Shiba state by scanning tunneling spectroscopy (STS), one needs
a relatively large gap. ${\rm MgB}_2$ is a perfect candidate in
this respect since it is a conventional superconductor  that has
an unusually high critical temperature, $T_c = 39\, {\rm K}$
\cite{Akimitsu}. Second,  ${\rm MgB}_2$ has a hexagonal
$AlB_2$-type structure and a highly anisotropic band structure
\cite{Budko,Mazin}. As we shall see below,  this leads to a clear
separation of the multiple Shiba states. Finally, there is
increasing   experimental evidence from spectroscopic
\cite{Giubileo,Szabo,JohnZ}, photoemmision \cite{Tsuda} and
transport \cite{Wang} measurements that favor the presence of two
different superconducting gaps in ${\rm MgB}_2$. It is therefore
an interesting question, how  the presence of these two gaps
influences the structure of Shiba states. 
As mentioned above, ${\rm MgB}_2$ crystallizes in the hexagonal
${\rm AlB}_2$-type structure \cite{Budko} in which the $B^{-}$
ions constitute graphite-like sheets in the form of honeycomb
lattices separated by hexagonal layers of ${\rm Mg}$ ions. Band
structure calculations \cite{Mazin} indicate that ${\rm Mg}$ is
substantially ionized, and the bands at the Fermi level derive
mainly form Boron $p$ orbitals. Four of the six $p$ bands cross
the Fermi energy, and the Fermi surface consists of quasi-$2D$
cylindrical sheets, due to ${\rm B}$ - $ p_{x,y}$ orbitals, and a
$3D$ tubular network (mostly originating from ${\rm B}$ - $p_z$
orbitals). It is believed that both structures participate in the
formation of the superconducting state, though the gap  is very
different on the tubular network and on the cylindrical sheets.

\begin{figure}[h]
\centerline{\includegraphics[width=2in]{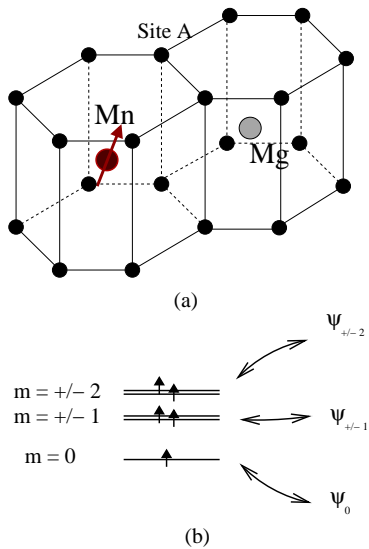}} \vspace*{3ex}
\caption{\label{fig:cage}
Fig.~a: Local environment of an ${\rm Mn}$ ion in ${\rm MgB}_2$. The ${\rm Mn}$ ion is located in a hexagonal cage.
Fig.~b: Level structure and crystal field splitting of the ${\rm Mn}^{2+}$ core states.
}
\end{figure}

We begin the quantitative analysis of the magnetic impurity
problem by establishing how the ${\rm Mn}$ spins couple to the
conduction electrons. The interaction part of the Hamiltonian
depends on the specific location and electronic structure of the
magnetic impurity considered. In what follows, we provide a
detailed analysis for ${\rm Mn}$ impurities, which presumably substitute
the ${\rm Mg}$ ions, and take an ${\rm Mn}^{2+}$ configuration
with a half-filled $d$-shell and a spin $S=5/2$ \cite{Mndoping},
though similar considerations hold for other types and positions
of magnetic impurities.
 As shown in Fig.~\ref{fig:cage}, the five-fold degeneracy of the $d$-states is
lifted by the local hexagonal crystal field
into three multiplets that we can label by the original angular momentum
quantum numbers $m$  of the d-states $|m\rangle$. Each of these states
is occupied by a single electron, and hybridizes through
a hybridization $V_m$ with  a specific local combination of $p$-states,
$\psi_m$, that we constructed
explicitly in the framework of a simple tight binding description of the
$p$-bands \cite{future}.  The $\psi_m$'s transform the
same way under the hexagonal point group as the $|m\rangle$'s.
Incorporating the local Hund's rule coupling and Coulomb interaction
and the $d$-states and  integrating out virtual charge fluctuations to them
we arrive at the following  exchange Hamiltonian:
\begin{equation}
H_{\rm int} = \frac 12 \sum_{m=-2}^2 J_m \;{\mathbf S} \;(\psi^\dagger_m
{\boldsymbol \sigma}
\psi_m)\;,
\label{eq:H_int}
\end{equation}
where the exchange couplings satisfy $J_m= J_{-m} \sim V_m^2/\Delta E$,
with $\Delta E$ the characteristic energy of charge fluctuations,
and the $\sigma$'s denote the Pauli matrices.
Note that Eq.~(\ref{eq:H_int}) must conserve $m$ by symmetry, and
therefore {\em five} independent channels couple simultaneously
to the impurity. This is the ultimate reason for the appearance of
multiple Shiba states.

To analyze the Shiba states in detail, we described
the superconducting state of  ${\rm MgB}_2$ by the following
non-interacting BCS Hamiltonian:
\[
H_0 =  \sum\limits_{\alpha, {\bf k},\sigma }
\varepsilon_{{\bf k},\alpha} c_{{\bf
k} \alpha \sigma }^{\dagger}c_{{\bf k} \alpha \sigma }
+ \sum\limits_{\alpha, {\bf k}} \Delta _\alpha
\bigl( c_{-{\bf k} \alpha \downarrow
}^{\dagger} c_{{\bf k} \alpha  \uparrow }^{\dagger}
+ {\rm h.c.}
\bigr).
\]
Here $c_{ {\bf k}\alpha\sigma }^{\dagger}$ creates an electron
with momentum ${\bf k}$, spin $\sigma$, energy
$\varepsilon_{{\bf k},\alpha}$, and wave function $\phi_{\alpha,\bf k}$
in one of the four bands that cross the Fermi energy ($\alpha=1\dots4$).
The kinetic energies
$\varepsilon_{{\bf k},\alpha}$ and the corresponding wave functions have
been constructed using a simple tight binding model, which has proved
quite accurate in the vicinity of the Fermi surface \cite{future}.
We assume further that the superconducting gaps
take two different values in the $p_{x,y}$ and $p_z$ bands
$ \Delta_{xy}\approx 7.5$ meV and $\Delta _z\approx 2.5$ meV.

As a next step, we  express the interaction part
Eq.~(\ref{eq:H_int}) in terms of the operators $c_{ {\bf
k}\alpha\sigma }^{\dagger}$ and  re-express the Hamiltonian in
terms of Nambu spinors \cite{Nambu}
$\Phi_{{\bf k },\alpha} \equiv
\{ c_{{\bf k},\alpha,\uparrow}, c_{{\bf k},\alpha,\downarrow},
 -c^\dagger_{-{\bf k},\alpha,\downarrow},
c^\dagger_{-{\bf k},\alpha,\uparrow} \}$ to obtain
\begin{eqnarray}
H & = & \sum_{{\bf k},\alpha}
\Phi_{{\bf k },\alpha}^\dagger (\varepsilon_{{\bf k},\alpha} \tau^z +
\Delta _\alpha \tau^x)\Phi_{{\bf k },\alpha}
\label{eq:nambu}
 \\
&+&
\sum\limits_{\scriptstyle {\bf k},{\bf k}^{\prime }\atop \scriptstyle
\alpha,\beta,m} \frac12
  J^{\alpha \beta}_m \; f^*_{m,\alpha}({\bf k})f_{m,\beta}({\bf k}^\prime)
\;\Phi_{{\bf k},\alpha}^{\dagger} \;{\boldsymbol \sigma}\cdot{\bf S}\;
\Phi_{{\bf k}^{\prime },\beta }\;
\;.\nonumber
\end{eqnarray}
The $\tau^i$'s here denote Pauli matrices acting in the
pseudospin (charge) index of the Nambu spinor. In
course of the derivation we made use of time reversal symmetry
and doubled the Hilbert space so that
the components of the Nambu spinors in Eq.~\ref{eq:nambu}
must be considered as independent variables.
The form   factors  $f_{m,\alpha}({\bf k})$
are normalized such that they  satisfy the
orthogonality  relation on the  Fermi surface $S_\alpha$:
\begin{equation}
\int\limits_{S_\alpha} d^2{\bf k} f_{m,\alpha}({\bf k})
{f^*_{m',\alpha}({\bf k})} = \delta_{m,m'} S_\alpha\;.
\label{eq:orthogonal}
\end{equation}
We determined both the couplings $J^{\alpha \beta}_m$ and
the  form factors  within the tight binding model of ${\rm Mg B}_2$
\cite{future}: In this approximation  -- apart from an overall prefactor
in  $J^{\alpha \beta}_m$ --  they uniquely depend on
the crystal structure of ${\rm Mg B}_2$.
The couplings   $J^{\alpha \beta}_m $ in Eq.~\ref{eq:nambu}
approximately satisfy
 $J^{\alpha \beta}_m \approx (J^{\alpha\alpha}_m J^{\beta\beta}_m)^{1/2}$
and are equal in channels $\pm m$ by symmetry.
In the rest of the paper, we neglect quantum fluctuations of  the ${\rm Mn}$ spin
and treat $S$ as a classical variable.
In this limit we can solve the impurity problem exactly.
To determine the STM spectrum  we first
compute the Nambu Green's function in the presence of the
impurity:
\begin{eqnarray}
&& G(\alpha,\beta,{\bf k},{\bf k}^{\prime }, \omega ) =
\delta_{{\bf k, k'}} \delta_{\alpha,\beta} G_\alpha^{(0)}({\bf k}, \omega )
\label{eq:full}
\\
& + & G_\alpha^{(0)}({\bf k}, \omega )
\sum\limits_{m} {1\over\Omega}\;
f^*_{m,\alpha}({\bf k})
T_{m}^{\alpha,\beta} ( \omega) \; f_{m,\beta}({\bf k}^\prime)
 G_\beta^{(0)}({\bf k}^{\prime }, \omega )\;,
\nonumber
\end{eqnarray}
where $ G^{(0)}_\alpha({\bf k}, \omega )=\left( \omega
 -\varepsilon _{ \alpha,{\bf k}}\tau^z- \Delta_\alpha \tau^x \right) ^{-1}$
denotes the non-interacting momentum space Nambu Green's function
in band $\alpha$. Eq.~(\ref{eq:orthogonal}) guarantees that  the quantum
number $m$ is   conserved and thus the T-matrix $T^{(m)}$ can be computed
independently for  each channel $m$. The computation simplifies further
because  $ G^{(0)}_\alpha$
 depends on ${\bf k}$  only through the energy $\varepsilon_{\alpha,\bf k}$.
 As a result, we can express $T_m^{\alpha,\beta}$ simply as
 \begin{equation}
 {\bf T}_{m}( \omega )= {\bf J}_{m} {\bf S \;\sigma} /2 \;
 \left[1 -  {\bf F}(\omega ) {\bf J}_{m} {\bf S \;\sigma} /2\right]^{-1}\;,
\label{eq:T}
 \end{equation}
 where we introduced a matrix notation in the band indices,
$J_{m}^{\alpha\beta}
 \to {\bf J}_{m}$, $T_{m}^{\alpha,\beta} \to {\bf T}_m$.
The matrix $F_{\alpha,\beta}( \omega ) =
\delta_{\alpha,\beta}F_{\alpha}(\omega ) \to {\bf F}$
in Eq.~\ref{eq:T} simply denotes the local Nambu propagator:
 \begin{equation}
 F_{\alpha}(\omega ) =
 \int\limits_{-\infty}^\infty  \; d\varepsilon \; \varrho_\alpha
 {1\over \omega - \varepsilon \tau^z - \Delta_\alpha  \tau^x}\;,
 \end{equation}
 with $\varrho_\alpha$ the density of states at the Fermi energy in band
 $\alpha$. For $|\omega|< \Delta_\alpha$ the function
$ F_{\alpha}(\omega )$ has only real parts and simplifies to
   $F_\alpha (\omega )=-\pi \varrho_\alpha
   (\omega +\Delta_\alpha \tau^x)/ (\Delta _\alpha^2-\omega ^2)^{-1/2}$,
   while for $|\omega|> \Delta_\alpha$ it is purely imaginary.

Impurity bound states and resonances can be identified from the pole structure
of the T-matrix: True bound states correspond to zeros of the determinants
${\rm det}\{{\bf T}_m^{-1}(\omega)\}$ on the real axis, and must
satisfy $|\omega| < \Delta_\alpha$ for all bands. Zeros in the
vicinity of the real axis, on the other hand, correspond to resonances.
We found that each of the five channels generates a pair of
bound states, two of which are  doubly degenerate by symmetry.

Let us first discuss the case of two bands only. There the
calculations further simplify and the energies  $E$ of the
bound states  satisfy the  following implicit equations:
\begin{eqnarray}
& (1 - g_{m}^{11} \alpha_1(\pm E)) (1-  g_{m}^{22} \alpha_2(\pm E )) =
\nonumber \\
& \phantom{nnn} (g_{m}^{12})^2 \alpha_1 (\pm E )
\alpha_2(\pm E)\;,
\label{eq:implicit}
\end{eqnarray}
with $g_m^{\alpha\beta} \equiv \pi S \sqrt{\varrho_\alpha \varrho_\beta}
J_m^{\alpha\beta} /2$ the dimensionless couplings in channel $m$,
and
$$
\alpha_\alpha(\omega) = \left({\Delta_\alpha + \omega
\over \Delta_\alpha - \omega}\right)^{1/2}\;.
$$
In the absence of inter-band coupling, $g^{12}=0$
Eq.~(\ref{eq:implicit}) would have two independent solutions for
each $m$, and the presence of the second superconducting gap would
double the number of Shiba states.  For small $g^{12}$'s the bound
states with energies $\Delta_{z}< |E| < \Delta_{xy}$ evolve into
resonances due to mixing with the continuum. We have in fact shown
that in reality $g^{12}$ is large. Consequently, the naive
expectation mentioned above - based solely on the straightforward
extension of Shiba's result for two bands - is not correct. Using
the approximate relation for $(g_{m}^{12})^2 \approx
g_{m}^{11}g_{m}^{22}$, we obtain that
\begin{equation}
g_{m}^{11} \alpha_1(\pm E) + g_{m}^{22} \alpha_2(\pm E) = 1\;,
\end{equation}
which has only one pair of solutions for each $m$. Exchange
coupling between the two bands thus removes half of the
resonances, and the number of Shiba states is not increased due to
the presence of the second gap in this case. We thus obtain
altogether {\em five} pairs of Shiba states corresponding to the
five channels, but two of them are two-fold degenerate because of
the symmetry $J^{\alpha\beta}_m =  J^{\alpha\beta}_{-m}$.

We are most interested in the total and spin resolved
tunneling density of states in the vicinity of the ${\rm Mn}$ sites, which
are directly measured by conventional and spin polarized STM
techniques\cite{spinpolSTM}, respectively, and are related  to  the
local Nambu Green's function $G({\bf R},p_\alpha, \omega)$
at site ${\rm R}$ and  orbital $p_\alpha$ ($\alpha=x,y,z$)
as
\begin{eqnarray}
\varrho_{c,\alpha}(\omega) &=& - {1\over 2\pi} {\rm Im} {\rm Tr}
\left \{ G({\bf R},p_\alpha, \omega) {1 + \tau_z \over 2} \right
\}\;,
\\
\varrho_{s,\alpha}(\omega) &=& - {1\over 2\pi} {\rm  Im} {\rm Tr}
\left \{ G({\bf R},p_\alpha, \omega) \;{\boldsymbol \sigma }
{\mathbf n}\; {1 + \tau_z \over 2} \right \}\;,
\end{eqnarray}
with $\mathbf n$ a unit vector pointing in the direction of the ${\rm Mn}$ spin.
These can be computed easily from (\ref{eq:full}) in terms
of the electronic wave functions  $\phi_{\alpha,\bf k}$ \cite{future}.

To obtain a quantitative estimate for the STM spectra we performed
a lengthy, but straigthforward tight-binding calculation to
determine numerically the form factors, the exchange couplings and
the electronic wave functions above in various geometries. In
particular,  we generalized the discussions above for  the case of
a semi-infinite half plane and computed the STM spectra as a
function of the ${\rm Mn}$ positions \cite{future}. 
In these calculations we
assumed that the value of the order parameter does not depend on
the distance from the surface.

\begin{figure}[h]
\centerline{\includegraphics[width=3.2in]{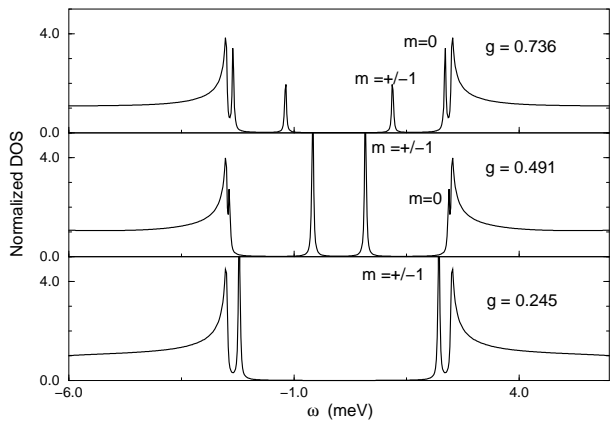}} \vspace*{3ex}
\caption{\label{fig:positionA} Normalized density of states for
site A  as function of frequency for different values of $g$. }
\end{figure}

\begin{figure}[h]
\centerline{\includegraphics[width=3.2in]{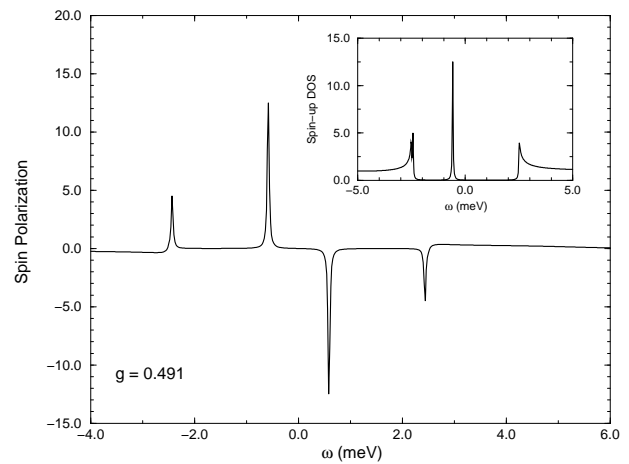}} \vspace*{3ex}
\caption{\label{fig:spin}
Spin polarization and spin resolved density of states (inset) for site A and $g=0.491$.
}
\end{figure}

In Fig.~\ref{fig:positionA}  we show the
density of states for the case  of a half-infinite sample for
various values of the dimensionless exchange coupling, $g \equiv
\frac 15 \sum_{m,\alpha} \varrho_{\alpha} J_m^{\alpha \alpha}$, with
$\varrho_\alpha$ the  density of states in band $\alpha$ per one
spin direction at the Fermi energy. We assumed  that ${\rm MgB}_2$
is cleaved along a Boron sheet and that the ${\rm Mn}$ ion sits
right below this Boron layer. In the STM experiment one presumably
tunnels into the $p_z$ orbitals  oriented perpendicularly to the
Boron plane. Therefore we plotted the  DOS that  corresponds to a
local $p_z$ orbital at position A.  
For generic values of the exchange coupling
usually two well-separated pairs of resonances can be observed,
corresponding to the $m=\pm1$ and $m=0$ channels. The exchange
couplings in channels $m=\pm 2$ are much smaller than
those in channels  $m=\pm 1$ and $m=0$, therefore the bound
state appears  at the superconducting coherence peak.
The wave functions of the Shiba states and thus the
amplitudes of the  corresponding resonances in the spectrum depend
a lot on the tunneling position\cite{future}. Moreover, as we show in
Fig.~\ref{fig:spin}, the Shiba states are strongly spin-polarized.
Therefore,  even if a peak happens to be close to the BCS
coherence peaks, a spin-polarized STM can distinguish the Shiba
state from the coherence peak. Thus, spin-polarized STM would
provide a perfect tool to identify these multiple Shiba states.

The presence  of spin-orbit coupling should not influence these results:
Since the $\pm m$ $d$-levels form time reversed pairs, the symmetry
$J_m = J_{-m}$ should hold even if we take spin-orbit coupling into account.
We conjecture  therefore that the Shiba states remain degenerate even in the
presence of spin-orbit coupling.

Finally, let us shortly discuss how these results generalize for 
 other compounds. 
Possible crystal field structures in other cases
and the corresponding exchange  Hamiltonians have been discussed for
various point symmetries in the seminal paper of Nozi\`eres and Blandin \cite{blandin}.
Though the form of the exchange Hamiltonian depends a lot
on the specific material, in many cases, similar to ${\rm Mn}$-doped ${\rm MgB}_2$,
several channels of conduction electrons couple to the local impurity
degrees of freedom, and result in multiple Shiba states. Thus the appearance
of multiple Shiba states seems to be a rather general phenomenon.

In conclusion, we have shown that  a magnetic impurity generally induces multiple 
Shiba states in the electronic structure of  ${\rm MgB}_2$.  
In particular for ${\rm Mn}$ we found five pairs of Shiba states in the gap, two
of which were two-fold degenerate. We computed both conventional
and spin-resolved  STM spectra near the impurity site and showed
that these states can be  clearly resolved by both methods.
Similar multiple Shiba states should appear in other superconductors due to the 
internal structure of the magnetic impurity.

{\em Acknowledgments} We are grateful to Drs. Dan Agterberg,
George Crabtree, J.C. Seamus Davis, 
Maria Iavarone, Goran Karapetrov, Igor Mazin,
Kaori Tanaka, Ali Yazdani,
and John Zasadzinski for useful discussions. This
work was supported by the U.S. Dept. of Energy, Office of Science,
under Contract No.~W-31-109-ENG-38, Hungarian Grants No.
OTKA F030041,  
T038162, 
the Bolyai foundation,
and the EU  ``Spintronics'' RTN.
BJ was supported by NSF-NIRT award DMR02-10519 and the Alfred P.
Sloan Foundation. ED was supported by US NSF grant DMR-0132874.

\end{document}